\documentclass[runningheads]{llncs}
\usepackage[T1]{fontenc}

\usepackage{booktabs}
\usepackage{amsmath}
\usepackage{graphicx}
\usepackage{subcaption}
\begin{document}
\title{Warning About AI Fallibility Increases Help-Seeking in an Intelligent Tutoring System}
\titlerunning{Warning About AI Fallibility Increases Help-Seeking}
%
\author{Tomohiro Nagashima\orcidID{0000-0003-2489-5016} \and
Mirella Hladký\orcidID{0000-0001-5565-3998}\and
Vera Rief\orcidID{0009-0001-5787-3748}}

\authorrunning{T. Nagashima et al.}
%
\institute{Saarland University, Saarland Informatics Campus, Saarbrücken, Germany \email{nagashima@cs.uni-saarland.de, mirella.hladky@uni-saarland.de, veri00001@stud.uni-saarland.de}}
\maketitle              
\begin{abstract}
Recent work in Technology-Enhanced Learning and Human-Computer Interaction highlights the importance of transparency and trust calibration in AI-supported learning environments as they pose a risk of hallucinations. In this study, we investigate whether a simple transparency intervention that warns students that a pedagogical agent may make mistakes affects learner behavior in a math intelligent tutoring system. We conducted a classroom experiment with 252 school students using two system versions: one including a warning message about potential system errors, and one that does not mention potential errors. Using log data, we analyzed students’ problem-solving performance data, including help-seeking behavior, error rate, and time-on-task. Results show that students who were warned about potential AI errors requested significantly more hints than those in the other condition, even though the actual system behavior was exactly the same. This finding suggests that lightweight transparency interventions can influence learners’ interaction strategies without necessarily improving or impairing immediate performance.

\end{abstract}
\section{Introduction}
AI-based tutoring systems are increasingly used to support learners. Decades of research on Intelligent Tutoring Systems (ITSs) have demonstrated that adaptive support and scaffolding, such as feedback, hints, and cognitive, metacognitive, and affective guidance, can effectively enhance student learning across domains \cite{kulik2016effectiveness}.

More recently, researchers and developers have begun integrating Large Language Models (LLMs) into tutoring systems to enable more flexible, personalized, and conversational forms of support \cite{meyer2024using}. While these models offer new opportunities for scalable and adaptive instruction, their use in educational settings introduces important challenges. In particular, LLMs are prone to generating incorrect or misleading information, often referred to as hallucinations \cite{gupta2025beyond,hendrycks2021measuring}. In the context of tutoring, where accuracy is critical, such errors may negatively affect student understanding if they go unnoticed.

At the same time, prior research suggests that exposure to incorrect information is not necessarily detrimental to learning. Studies on erroneous worked examples have shown that engaging with and correcting incorrect solutions can foster deeper conceptual understanding compared to studying correct solutions alone \cite{booth2013using}. Similarly, recent work in LLM-based tutoring suggests that even highly inaccurate or hallucinated feedback can, under certain conditions, support learning gains \cite{steinbach2025llms}. These findings point to a more nuanced view in which errors themselves are not inherently harmful but their effectiveness rather depends on how learners engage with them.

This raises an important question for the design of AI-based tutoring systems: \textit{How can learners be supported in engaging critically with potentially fallible system output?} One promising approach is to make system uncertainty or fallibility explicit through transparency cues, such as a warning that the system may make mistakes. Such cues (e.g., “ChatGPT can make mistakes”) may encourage learners to become more metacognitive, which can potentially increase self-regulated behaviors such as verification, reflection, and help-seeking. However, prior work has not provided empirical evidence on the effectiveness of such lightweight interventions in authentic classroom settings.

In this paper, we investigate the effects of a simple transparency cue, a warning message indicating that the pedagogical agent may make mistakes, on student behavior in a classroom-based math tutoring system. We ask the following RQ: \textit{Does a warning message indicating that an AI-based tutoring system can make mistake influence students' problem-solving behaviors (errors, help-seeking, time spent)?}

\section{Methods}

\subsection{Participants}
We conducted a classroom experiment at a secondary school in Tokyo, Japan. Participants included 270 seventh-grade students (aged 12 - 13) in seven classes taught by one teacher. The study was conducted in Fall 2025. 

\subsection{Materials}
\begin{figure}[htpb]
    \centering
     \begin{subfigure}[t]{0.48\textwidth}
        \includegraphics[width=\linewidth]{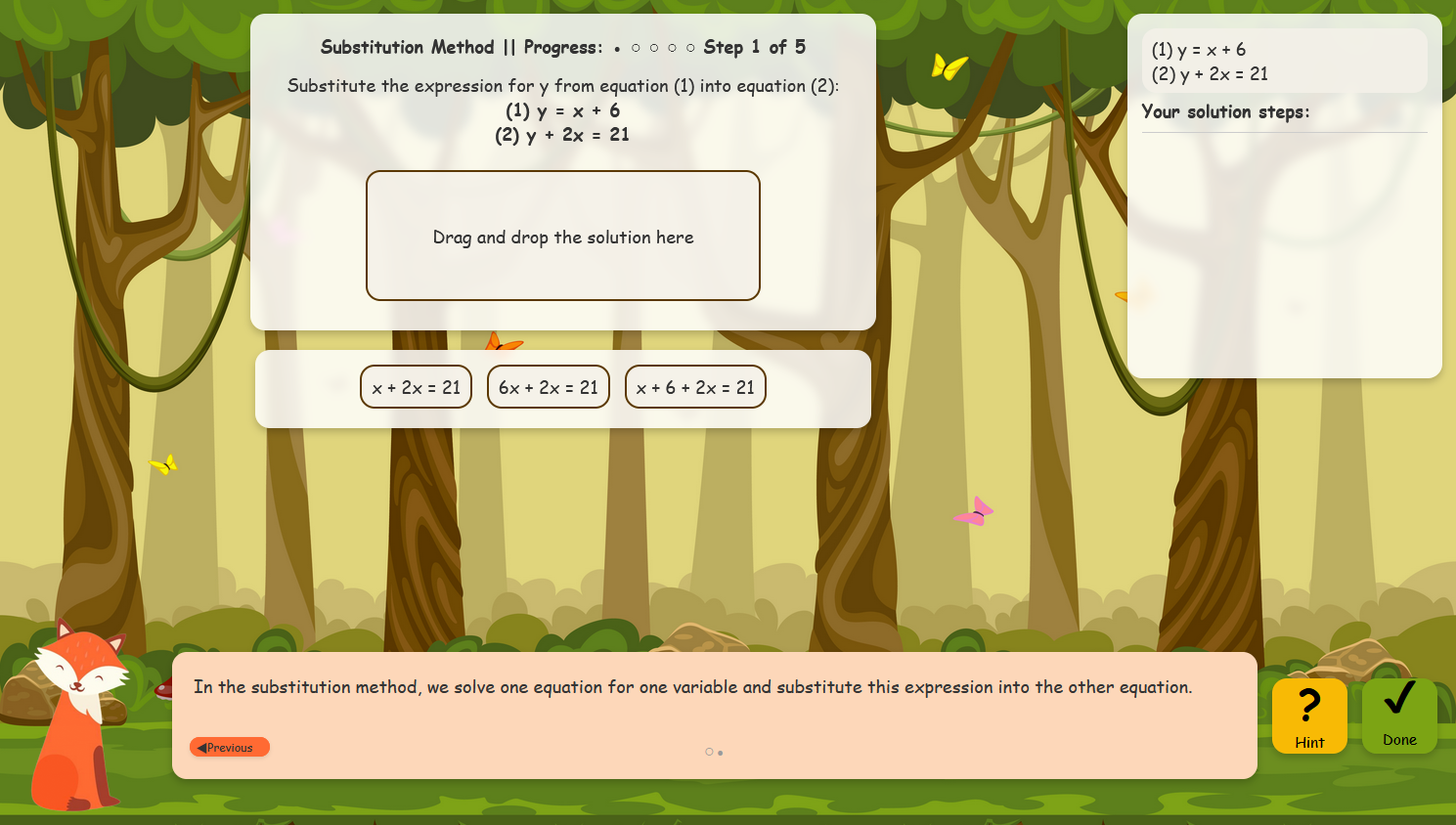}
        \caption{Drag-and-drop action}
        \label{fig:drag}
      \end{subfigure}
      \hfill
      \begin{subfigure}[t]{0.48\textwidth}       
        \includegraphics[width=\linewidth]{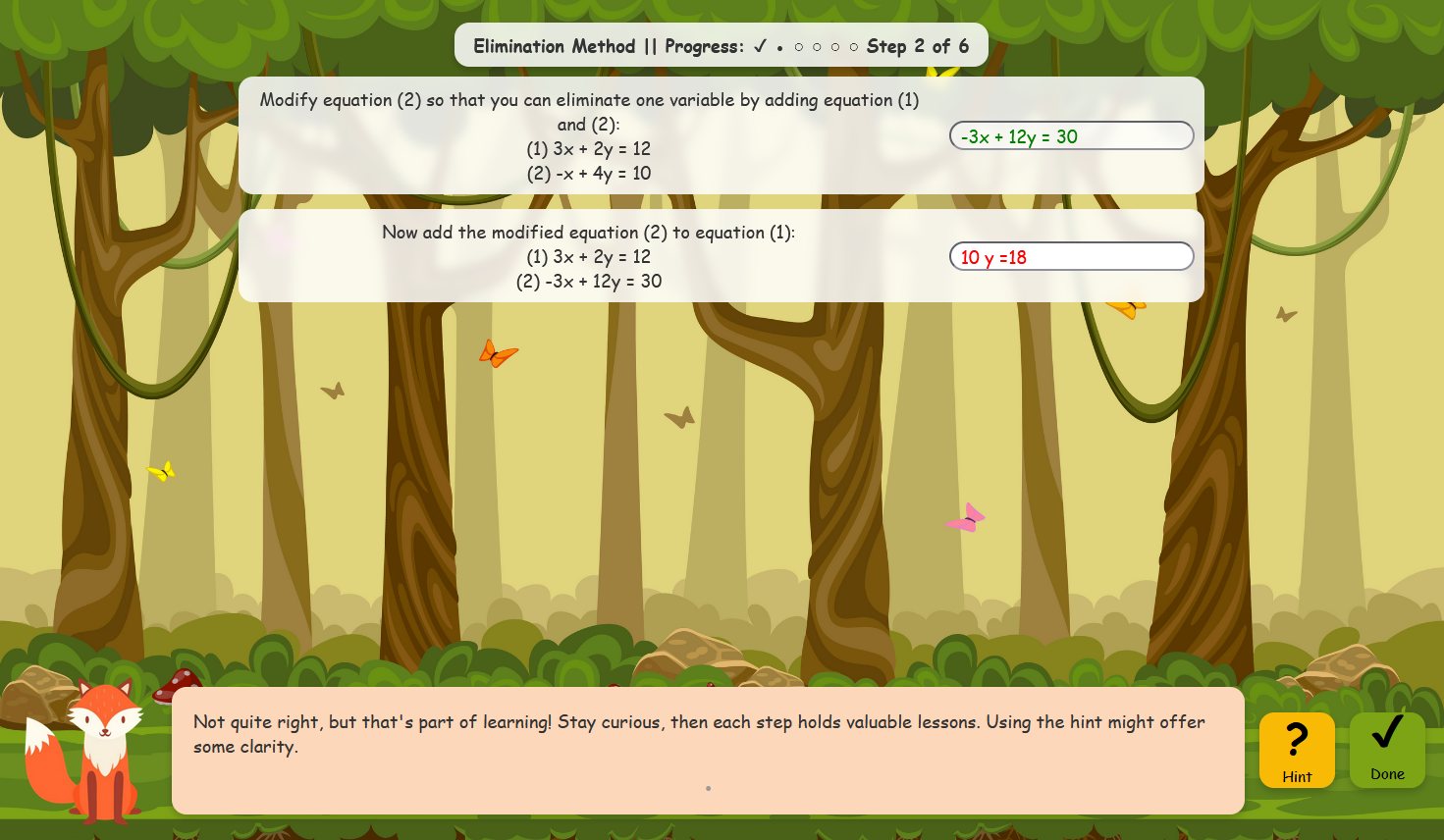}
        \caption{Free-text input mode}
        \label{fig:freeInput}
      \end{subfigure}
  \caption{Interactions modes of the tutor. (a) Students initially engage with scaffolded drag-and-drop interactions, which will be replaced later with (b) free-text input mode to practice step-by-step problem solving}
  \label{fig:interactions}
\end{figure}

We developed two versions of a math intelligent tutoring system, developed with the Cognitive Tutor Authoring Tools (CTAT) \cite{aleven2006cognitive}. The tutor is designed to support secondary school students in practicing systems of linear equations through two complementary problem-solving modalities. In the initial stages, students interact via drag-and-drop manipulations, selecting the next strategic step from a set of predefined options to scaffold their understanding. As they progress, the tutor transitions to a free-text input mode, allowing students to enter mathematical expressions independently and engage in more open-ended problem solving (see Figure \ref{fig:interactions}).

\setlength{\parskip}{0pt}

Throughout the problem sets, students are guided by a fox-shaped pedagogical agent (called “Maddox”) that provides hints and targeted feedback. To ensure accuracy and consistency, these hints and feedback messages are rule-based and hard-coded, without reliance on LLM-generated responses (i.e. there are no “errors” in system-provided hints and feedback).

\begin{figure}[htpb]
    \centering
     \begin{subfigure}[t]{0.48\textwidth}
        \includegraphics[width=\linewidth]{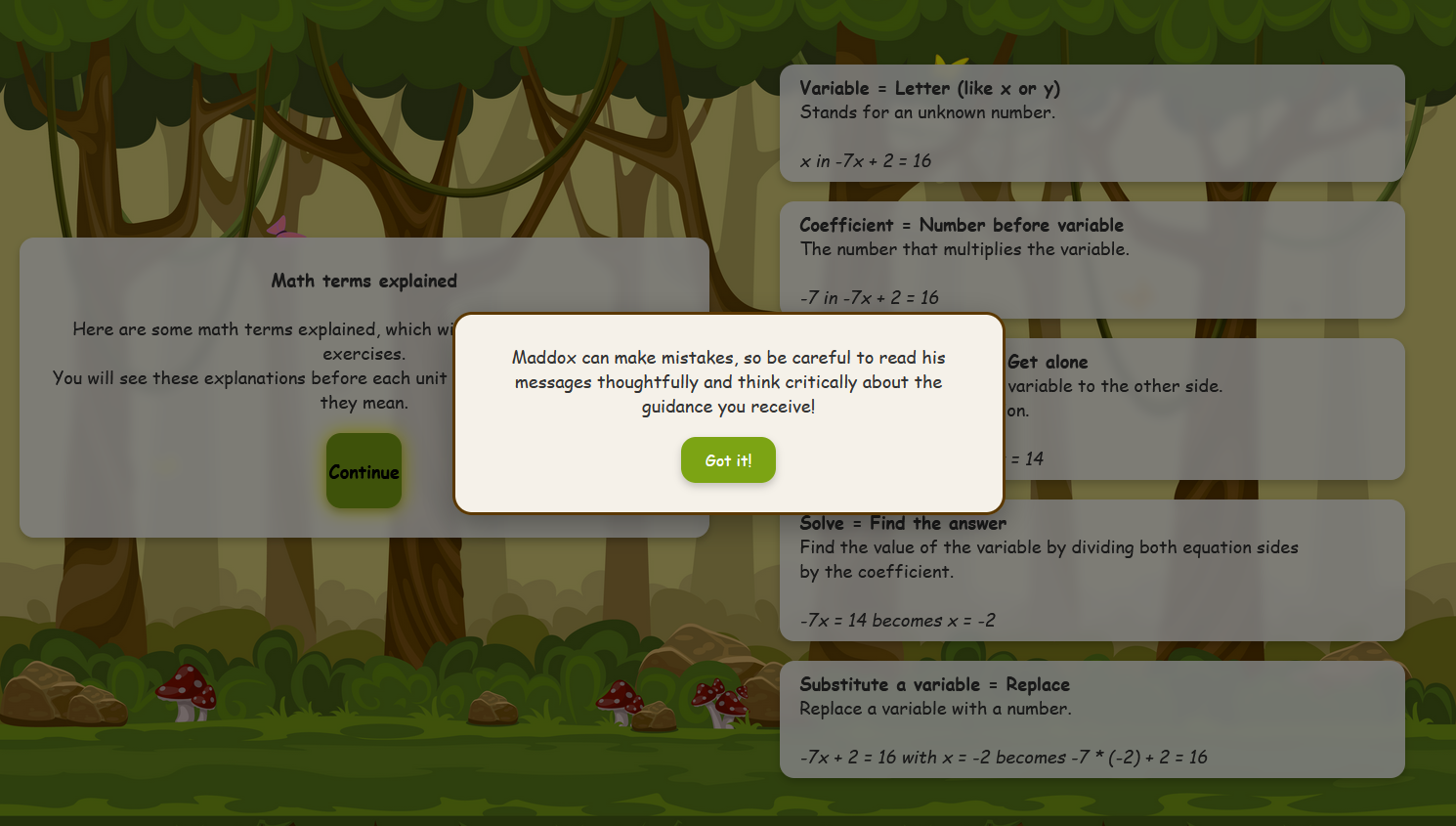}
        \caption{Mistake popup: Mistake condition}
        \label{fig:mistake}
      \end{subfigure}
      \hfill
      \begin{subfigure}[t]{0.48\textwidth}       
        \includegraphics[width=\linewidth]{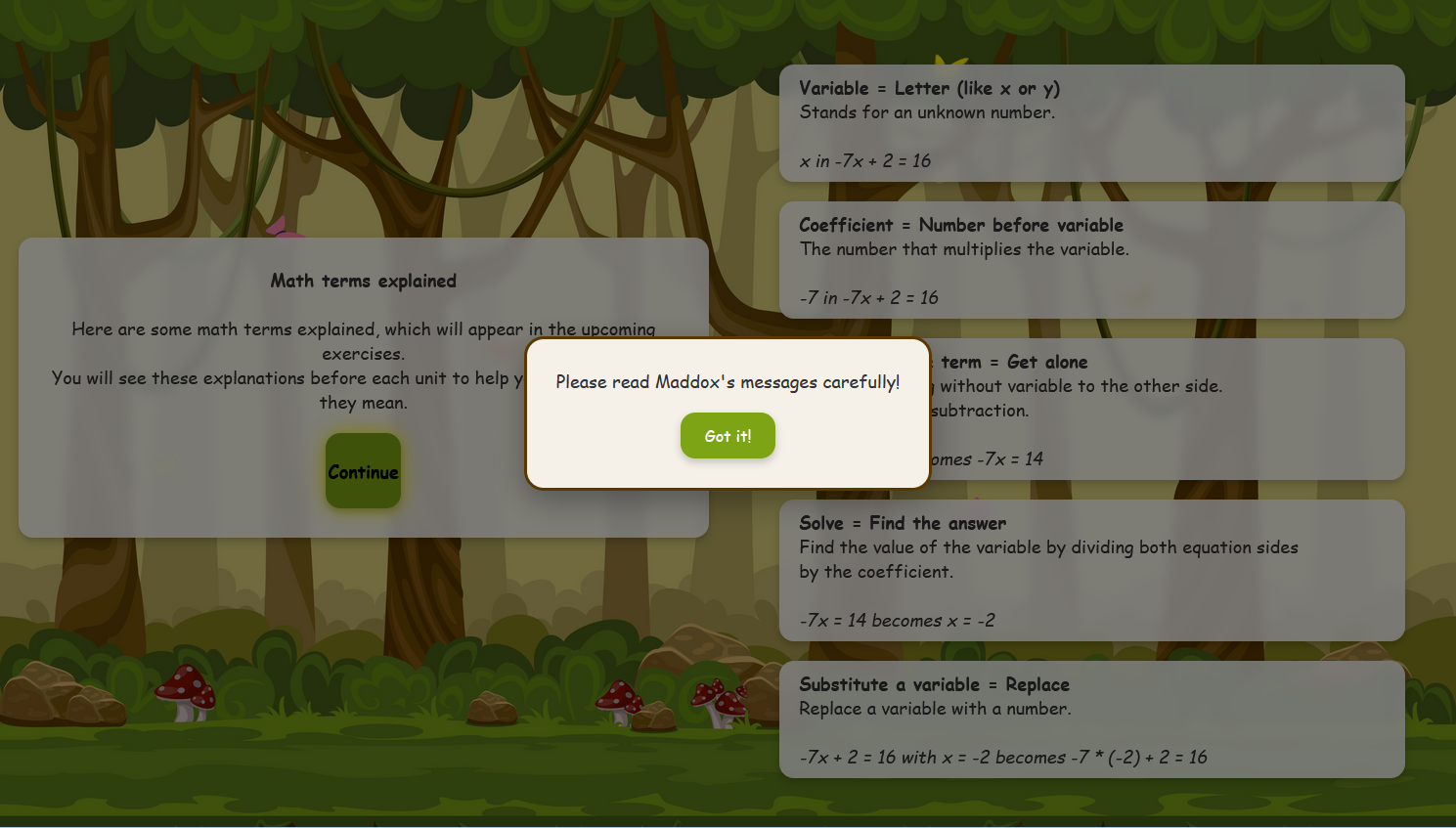}
        \caption{Normal popup: NoMistake condition}
        \label{fig:normal}
      \end{subfigure}
  \caption{Popup warning with a message}
  \label{fig:popups}
\end{figure}

The two versions we compared in the study only differed on whether students received a message that indicates that the pedagogical agent can make mistakes (Mistake condition) or not (NoMistake condition). The message is displayed as a popup window on the introduction page (see Figure \ref{fig:popups}), which appears three times throughout the tutor learning, when students are introduced a new solution method (students practiced three different solution methods to identifying two variables in the system of linear equations, called substitution, equalization, and elimination methods, using two different interface versions, Figure \ref{fig:interactions}). 

As seen in Figure \ref{fig:popups}, the Mistake condition received a message that says “Maddox can make mistakes, so be careful to read his messages thoughtfully and think critically about the guidance you receive!” while students in the NoMistake condition received a message “Please read Maddox's messages carefully!” The distinction focused on whether the warning message mentioned the possibility of hallucination or not, instead of, for example, giving no warning to students in the control condition (to isolate the effect of disclosing the \textit{possibility of making mistakes}, and not to test the general effect of providing a popup message with a warning that students should read Maddox's guidance carefully). Importantly, as hints and feedback messages were hard-coded without using LLMs, the two systems behaved exactly the same in the actual interactions, and no “errors”  were designed in the system. This decision draws on Human-Computer Interaction (HCI) research regarding the “placebo effect” of AI. This phenomenon demonstrates that users alter their expectations and behaviors based on perceived capabilities of AI, even when the underlying AI performance remains the same \cite{10.1145/3529225}.

\subsection{Procedure}
The study took place during two regular class periods (50 minutes each) of the “Information Technology and Data Science” course, which is held once a week for every class. Session 1 included an introduction of the study, pre-test on linear systems of equations (15 min), and students worked on the ITS for 15 minutes. Session 2 started with 15 minutes of continued problem-solving practice with the ITS, post-test on linear systems of equations (isormophic to pre-test), and a short debriefing discussion with students. 

Twelve students in each class (35-41 students) were randomly assigned to the Mistake condition (using the Mistake version), the remaining students were assigned to the NoMistake condition (using the NoMistake version). While group sizes were unequal due to a concurrent data collection focusing on the NoMistake version of the tutor, this did not affect the random assignment process. (the groups remained comparable across all other characteristics). Although the imbalance reduces statistical power, running statistical tests on the unbalanced groups still remains valid \cite{ELKINS202675}. The study was approved by the Ethical Review Board of the affiliated university prior to data collection.

\section{Results}
Eighteen students were absent on the days of data collection and were therefore excluded from the analysis. The final sample consisted of 252 students.

To address our RQ, we analyzed students' problem-solving performance; specifically, we examined students' average error rate (per problem-solving step), help-seeking behavior (the rate of hint requests divided by the number of problem-solving steps), and average time spent per problem-solving step. These represent common problem-solving performance metrics used in other ITS research \cite{nagashima2021using}. Table \ref{tab:logdata} shows means and standard deviations of these metrics for each condition. Pre-test and post-test data are not reported in this paper as they were many students who could not take the posttest due to unexpected class cancellations.

We then conducted a series of linear regression analyses. Specifically, we regressed students’ problem-solving performance (error rate, help-seeking, and time spent) on the condition. Results showed that students in the Mistake condition requested hints more frequently than those in the NoMistake condition ($\beta = -0.33$, $t(235) = -2.33$, $p = .02$). The groups did not differ significantly on the error rate and average time spent (error rate: $\beta = -0.02$, $t(235) = -0.51$, $p = .61$; time spent: $\beta = -0.23$, $t(235) = -0.86$, $p = .39$).

\begin{table}[h]
    \centering
    \caption{Students' Mean Values of Error Rates, Help-seeking Behavior, and Time Spent per Step (Standard Deviations in Parentheses)}
    \label{tab:logdata}
    \setlength{\tabcolsep}{12pt} 
    \begin{tabular}{l c c c}
        \toprule
        \textbf{Condition} & \textbf{Error rate} & \textbf{Help-seeking} & \textbf{Time spent per step}\\
        \midrule
        Mistake   & 0.49 (0.45) & 1.09 (1.26) & 1.60 (1.92) \\
        NoMistake & 0.46 (0.40) & 0.75 (0.88) & 1.36 (1.89) \\
        \bottomrule
    \end{tabular}
\end{table}

\section{Discussion and Conclusion}
The growing use of AI-driven interventions in learning technologies enables more flexible and adaptive support for learners. However, it also introduces significant risks in tutoring contexts, as AI systems, particularly those based on Large Language Models (LLMs), may generate hallucinations during interactions. To investigate this, we conducted an experiment with 252 seventh-grade students, examining whether a warning message indicating that the tutoring system may make mistakes influences students’ problem-solving behavior. Our findings show that, even though the system did not produce any errors or hallucinations, the presence of a simple warning message led to more frequent help-seeking behavior compared to students who were not informed of the system’s potential fallibility.

This finding suggests that a simple, lightweight warning indicating that the system may make mistakes can encourage learners to adopt a more cautious and potentially more metacognitive approach to problem solving. Although an increased frequency of help-seeking alone does not necessarily reflect metacognitive engagement, it is possible that students became more reflective about the guidance provided. For instance, students may have questioned the reliability of Maddox’s feedback and therefore requested additional hints to verify or clarify the information. Alternatively, the warning may have increased students’ awareness of their own uncertainty in problem solving and they wanted to seek more support when they felt less confident in their solutions; that is, the warning may have shifted responsibility to the learner (as they could no longer rely on the tutor's guidance as an authoritative source), and may have promoted students' monitoring around their uncertainty (and requested help to reduce their own uncertainty feeling).

We acknowledge limitations of this work. First, the ITS did not introduce any errors. Future studies could test a warning for system that actually present errors to examine how students would handle errors if they are informed about potential fallibility. Further, our study was conducted in a specific setting in Japan. Findings from this work therefore may not generalize across contexts.

Overall, our findings highlight that even minimal transparency cues about AI fallibility can meaningfully shape learners’ interaction strategies. This result suggests the importance of carefully designing how uncertainty is communicated in AI-supported learning environments.



%
\bibliographystyle{splncs04}
\bibliography{references}

\end{document}